# Non-Volatile Control of Valley Polarized Emission in 2D WSe$_2$-AlScN Heterostructures

Simrjit Singh[a,b], Kwan-Ho Kim[a], Kiyoung Jo[a], Pariasadat Musavigharavi[a,c,d], Bumho Kim[e], Jeffrey Zheng[c], Nicholas Trainor[f], Chen Chen[g], Joan M. Redwing[f,g], Eric A Stach[c], Roy H Olsson III[a], Deep Jariwala[a]

[a]Department of Electrical and Systems Engineering, University of Pennsylvania, Philadelphia, PA, USA.
[b]Department of Applied Physics and Science Education, Eindhoven University of Technology, Eindhoven, The Netherlands.
[c]Department of Materials Science and Engineering, University of Pennsylvania, Philadelphia, PA, USA.
[d] Department of Materials Science and Engineering, University of Central Florida, Orlando, FL, USA.
[e]Department of Physics and Astronomy, University of Pennsylvania, Philadelphia, PA, USA
[f]Department of Materials Science and Engineering, Pennsylvania State University, University Park, PA, USA.
[g]2D Crystal Consortium Materials Innovation Platform, Materials Research Institute, Pennsylvania State University, University Park, PA, USA.

**Abstract:** Achieving robust and electrically controlled valley polarization in monolayer transition metal dichalcogenides (ML-TMDs) is a frontier challenge for realistic valleytronic applications. Theoretical investigations show that integration of 2D materials with ferroelectrics is a promising strategy; however, its experimental demonstration has remained elusive. Here, we fabricate ferroelectric field-effect transistors using a ML-WSe$_2$ channel and a AlScN ferroelectric dielectric, and experimentally demonstrate efficient tuning as well as non-volatile control of valley polarization. We measured a large array of transistors and obtained a maximum valley polarization of ~27% at 80 K with stable retention up to 5400 secs. The enhancement in the valley polarization was ascribed to the efficient exciton-to-trion (X-T) conversion and its coupling with an out-of-plane electric field, *viz.* the quantum-confined Stark effect. This changes the valley depolarization pathway from strong exchange interactions to slow spin-flip intervalley scattering. Our research demonstrates a promising approach for achieving non-volatile control over valley polarization and suggests new design principles for practical valleytronic devices.

**Introduction**

Valleytronics is an emerging field whereby, the electron's valley degree of freedom (DoF), rather than its charge, is exploited to store and process binary information.[1,2] Two-dimensional transition metal dichalcogenides (2D-TMDs) possess two inequivalent valleys in momentum space at the K and K' points in the first Brillouin zone. This makes them particularly attractive for exploiting valleytronic phenomena.[2,3] This results in a strong spin-valley locking effect that gives rise to valley-dependent optical selection rules, enabling selective control of carriers in the K and K' valleys using circularly polarized light of different helicity.[4] Furthermore, a large momentum separation between two valleys is expected to leads to a slow decay of the valley information. Owing to these fascinating properties, a variety of valley-contrasting phenomena such as the valley selective circular dichroism,[5,6] the optical Stark effect,[7–9] the valley Hall effect,[10,11] the valley Zeeman effect,[12,13] etc. have been successfully demonstrated using various 2D-TMD systems. In spite of these significant advancements, the development of highly tunable valleytronic devices, analogous to solid-state microelectronics, is required to make valleytronics a viable commercial technology.

In valleytronic devices that rely upon optical manipulation of valleys, the degree of valley polarization and the lifetime of the valley polarized states are critical measures. The valley lifetime of excitons in monolayer TMDs (ML-TMDs) is typically quite low, owing to strong electron-hole exchange interactions, leading to fast valley depolarization.[14,15] Additionally, the typically volatile nature of optically initialized valley polarization severely degrades the fidelity of valley information. Alternative strategies, such as the magnetic proximity effect, have been used to obtain non-volatile valley splitting in ML-TMDs.[16–18] However, the magnitude of valley splitting obtained was quite low (in the range of 0.1-0.2 meVT$^{-1}$). Further, the requirement of a high magnetic field increases the complexity and energy consumption of devices. Thus, it is important to look for alternative ways to achieve static and robust valley polarization for practical applications.

Electrical control of valley DoF is highly desired for large-scale integration and energy-efficient tuning. The electrical control of valley pseudospin in ML-TMDs has not been well studied. Few reports use electrostatic gating in standard field effect transistors (FETs) to control valley polarization.[19–21] The gate-controlled exciton-to-trion (X-T) conversion has been shown to prolong the lifetime of valley-polarized charge carriers by suppressing the electron-hole exchange interaction strength and, consequently, the intervalley scattering. Traditional electrostatic gating uses a linear dielectric and relies on electric field adjustments and Fermi-level tuning. The potential of non-volatile electrical control of valley polarization using non-linear ferroelectric dielectrics has received limited investigation. This unexplored approach offers novel possibilities for valley manipulation and control.

Ferroelectric materials possess a spontaneous polarization, originating from their non-centrosymmetric structure, which remains robust even when the gate voltage is removed.[22] Moreover, the reversibility of ferroelectric polarization with the application of external electric fields could effectively enhance polarization-dependent functionalities of the valleytronic devices. Several reports using first-principles calculations indicate that ferroelectric polarization switching is a potential knob to modulate valley pseudospin in various 2D materials.[23–25] However, an experimental manifestation of this approach has yet to be demonstrated.

Herein, we demonstrate ferroelectric field effect transistors (FeFETs) using mechanically exfoliated as well as CVD-grown $WSe_2$ monolayers transferred on top of an AlScN dielectric that exhibit tunable, non-volatile, valley-polarized emission. AlScN is a recently discovered ferroelectric material displaying out-of-plane electrical polarization and superior ferroelectric properties, including very high remnant polarization (> 115 $\mu C/cm^2$) and coercive fields (2-6 MV/cm), compared to other commonly utilized ferroelectric materials.[26–32] Taking advantage of the ferroelectric properties of AlScN, we investigate valley polarization characteristics of ML-$WSe_2$ using ferroelectric polarization switching and demonstrate efficient tuning as well as non-volatile control of valley polarization. We measure a large array of FeFETs and observe a maximum degree of valley polarization (DVP) of ~27 % with stable retention up to 5400 secs at 80 K. The robust control of valley polarization was ascribed to the formation of charged excitons and their coupling with a large out-of-plane electric field associated with the large ferroelectric polarization of AlScN, leading to suppressed electron-hole exchange interactions and, thereby, suppressed intervalley scattering. Our results provide a novel integration strategy for non-volatile electrical control of valley properties in ML-TMDs for polarization tunable emitters.

**Results and Discussion:**

The prototypical FeFET in this study consists of monolayer $WSe_2$ as a semiconductor channel on top of 45 nm thick AlScN ferroelectric dielectric film grown on Pt(111)/Ti/$SiO_2$/Si substrate as . A schematic is shown as **Figure 1a**. The Ti/Au metal electrodes are used as the source and drain, and a conductive Pt layer are used as the bottom gate electrode. The channel length in the device is ~800 nm **(Figure 1b)**. The devices were fabricated using both mechanically exfoliated $WSe_2$ ($\mu m^2$ to $cm^2$ scale monolayers, using our modified exfoliation technique) as well as MOCVD-grown $WSe_2$ monolayers (**See Experimental Section for details**). A large array of the valleytronic transistors is also fabricated from both mechanically exfoliated and MOCVD-grown $cm^2$-scale $WSe_2$ monolayers. Optical images of representative devices are shown in **Figure 1c, 1d**. **Figures S1-S3,** in the **Supporting Information**, display photoluminescence (PL) intensity maps, atomic force microscopy (AFM) images, and Raman spectra for monolayer $WSe_2$ samples used in this study. The interfacial characteristics of a representative $WSe_2$/AlScN FeFET device are also imaged using bright field cross-sectional transmission electron microscopy (BFTEM), shown in **Figure 1e**. The high-angle annular dark field scanning transmission electron microscopy (STEM) image of the $WSe_2$/AlScN interface combined with elemental analysis (**Figure 1f-1i**) shows a monolayer of $WSe_2$ on top of AlScN. There is no evidence of a significant oxide layer between $WSe_2$ and AlScN.

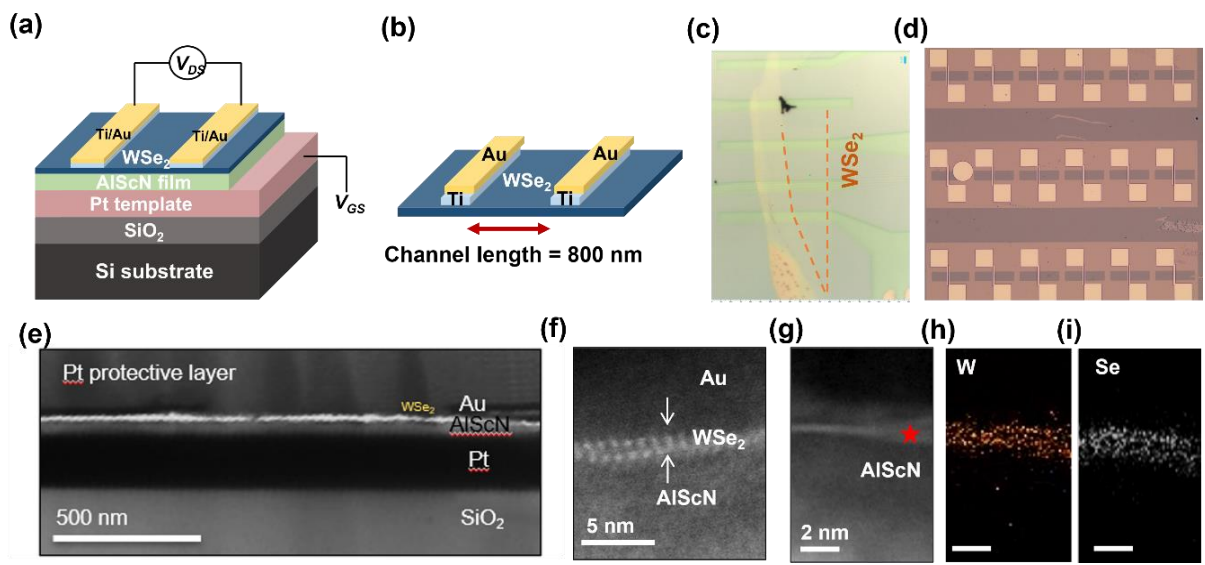

**Figure 1. The $WSe_2$/AlScN valleytronic device structure.** (a) and (b) Schematic representation of $WSe_2$/AlScN FeFET device. The Ti-Au metals are used as the source/drain electrodes and Pt layer is used as the bottom gate electrode. The channel length in the device is ~800 nm. (c) and (d) Optical images of the devices fabricated using mechanically exfoliated $WSe_2$ monolayers. A large array of the devices was fabricated from the large area mechanically exfoliated monolayers. (e) Cross-sectional BFTEM image of a representative $WSe_2$/AlScN device showing different material layers in the device. (f) Atomically resolved STEM image of the $WSe_2$/AlScN interface, which clearly shows a monolayer thickness of $WSe_2$. (g-i) Energy Dispersive X-Ray spectroscopic image of the W and Se in the monolayer $WSe_2$.

Ferroelectric polarization switching in the WSe$_2$/AlScN FeFET device is accomplished by the application of a gate bias ($V_g$). When a positive $V_g$ is applied, the electric dipoles in the ferroelectric-AlScN layer are polarized upward ($P_{up}$), pointing towards the channel (**Figure 2a**). As a result, the energy band alignment between AlScN and WSe$_2$ is tuned by the presence of the applied electric field. The conduction band of WSe$_2$ is bent below the Fermi level ($E_F$), resulting in the accumulation of mobile electrons in the WSe$_2$ channel (**Figure 2b**) and an accompanying increase in the channel conduction, as confirmed by the $I_d$-$V_d$ curves (**Figure 2e**) **(See linear scale I$_D$-V$_D$ characteristics of representative devices in Supporting Information Figure S11)**. A sharp increase in the drain current in the $P_{up}$ state confirmed the accumulation an excess of electrons in the WSe$_2$ channel. In contrast, when a negative '$V_g$' is applied, the AlScN is polarized downward ($P_{down}$) (**Figure 2c**). In this case, the valence band is bent above the $E_F$, and an excess of hole density is accumulated in the WSe$_2$ (**Figure 2d**), resulting in a decrease in the channel conductance (**Figure 2e**) **(See I$_D$-V$_G$ characteristics of representative devices in Supporting Information Figure S11)**. Importantly, the ferroelectric polarization maintains the channel state in the devices, even after the removal of the gate voltage, enabling non-volatile control of valley polarization. This is a conceptually interesting and novel form of optoelectronic memory that can store information both in the electrical conductance of the semiconducting channel of the FeFET as well as the degree of circular polarization of the photoluminescence. Specifically, the $P_{up}$ state can be described as the low resistance "*ON*" state and the $P_{down}$ state can be described as a high resistance "*OFF*" state. The presence of external charge carriers can strongly influence the PL characteristics of ML-TMDs: the photoexcited electrons and holes can combine with the extra charge carriers to form negatively or positively charged trions, even at room temperature. To determine the effect of ferroelectric polarization switching on the excitonic characteristics of WSe$_2$, we performed PL spectroscopy measurements on the devices when subjected to differently polarized states (**Figure 2f**). For ML-WSe$_2$ exfoliated onto a SiO$_2$/Si substrate, the PL spectra feature a highly intense emission from the neutral excitons ($X^o$) due to the large radiative recombination of photogenerated electrons and holes. In contrast, the PL spectrum from the WSe$_2$/AlScN device in the $P_{down}$ state was strongly suppressed, and red shifted to ~20 meV due to the formation of positively charged trions ($X_T^+$) in the hole-rich WSe$_2$. When the ferroelectric polarization direction is switched to the $P_{up}$ state, a high density of negative charges injected in WSe$_2$ results in the formation of negatively charged trions ($X_T^-$), leading to further suppression of the PL emission intensity and a red shift.[33,34] We observe that this PL modulation is quite reversible and can be switched to the "ON" and "OFF" states for multiple cycles of ferroelectric polarization switching (**Figure 2g**). To further verify the spatial homogeneity of the injected charge carriers with ferroelectric gating, we took PL intensity maps from the WSe$_2$/AlScN device in two oppositely polarized states **(Figure 2h-j and Figure S4)**. It is evident from **Figures 2i and 2j** that the PL intensity is altered significantly over the region close to the metal electrodes when the polarization is switched from the $P_{down}$ to the $P_{up}$ state, implying an efficient polarization switching and, importantly, that the switching is not limited to the area under the top metal electrodes only, but spreads laterally up to few µms away from the electrodes. The above features confirm ferroelectric polarization-induced gating can effectively modulate the optical characteristics of monolayer WSe$_2$ over the entire channel region.

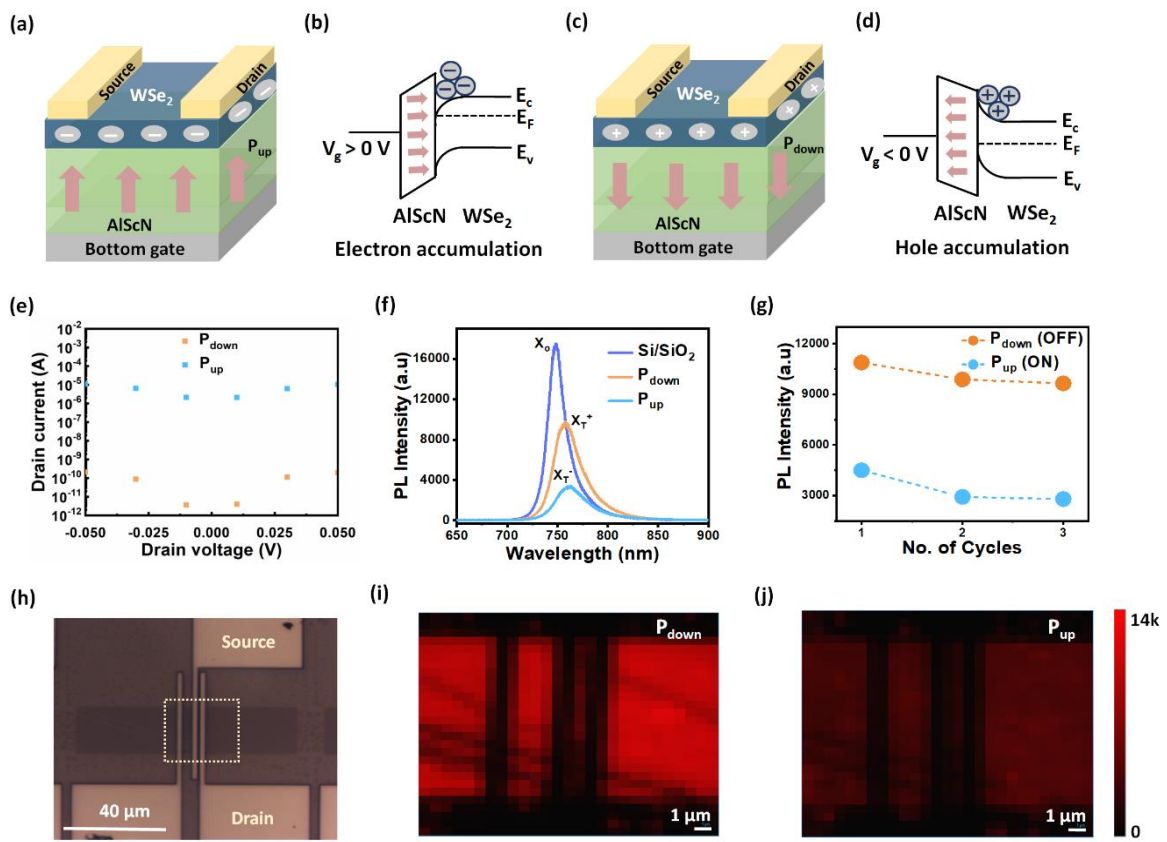

**Figure 2. Ferroelectric polarization switching in WSe$_2$/AlScN FeFET device, electrical and optical characteristics of the device.** (a-d) Schematic illustration of ferroelectric polarization switching i.e., (a) upward, $P_{up}$ and (c) downward, $P_{down}$ polarized states in the WSe$_2$/AlScN valleytronic device. (b and d) show the corresponding energy band alignment between the WSe$_2$ and AlScN in $P_{up}$ and $P_{down}$ states. (d) The positive gate bias switches the ferroelectric polarization upward, resulting in the accumulation of electrons in the WSe$_2$ channel (b). The negative gate bias switches the ferroelectric polarization downward, resulting in hole

accumulation in the WSe$_2$ channel (d). (e) The polarization switching is confirmed from the room temperature I$_d$-V$_d$ curves at V$_g$ of 0 V of the WSe$_2$/AlScN FeFET device in two oppositely polarized states. (f) PL spectra of WSe$_2$ transferred on Si/SiO$_2$, P$_{up}$ and P$_{down}$ polarized AlScN substrate. (g) Modulation of PL emission intensity in the P$_{up}$ and P$_{down}$ states for multiple cycles of the ferroelectric polarization switching. (h-j) The corresponding PL intensity maps of WSe$_2$/AlScN FeFET device in the P$_{up}$ and P$_{down}$ states shows significant quenching of PL emission in the P$_{up}$ state.

To obtain additional insight into the excitonic properties of WSe$_2$ resulting from ferroelectric gating, we performed temperature-dependent PL measurements. **Figures 3a-c** display temperature-dependent (295 to 80 K) PL spectra from the ML-WSe$_2$ samples exfoliated on both SiO$_2$/Si and AlScN substrates in two oppositely polarized states, under 633 nm laser excitation. At room temperature, the PL spectrum from the WSe$_2$/SiO$_2$/Si device displays emission from the neutral intralayer excitons ($X^o$), which exhibits a gradual blue shift of ~80 meV with significant PL linewidth narrowing upon cooling down to 80 K. This observation is in good agreement with previous reports from monolayer TMDs (**Figure 3a**).[35,36] At 80 K, a shoulder peak corresponding to the formation of trions is also observed due to unintentional doping in as-exfoliated monolayer WSe$_2$.[37] As shown in **Figure 3b**, the PL emission from the WSe$_2$/AlScN device polarized in the P$_{down}$ state is red shifted to ~20 meV below the $X^o$ emission energy and exhibits significant PL broadening, possibly due to the modified dielectric environment. This can be ascribed to the formation of positively charged bright trions ($X_T^+$).[38,39] As the temperature is decreased, the $X_T^+$ displays a blue shift of ~10 meV, which is significantly less than the shift observed for $X^o$ (~ 80 meV) in the WSe$_2$/SiO$_2$/Si device. Interestingly, at 80 K, a large red shift of ~110 meV, with significant linewidth broadening in the $X_T^+$ PL peak relative to the PL peak position of neutral excitons was observed. The red shift in the PL peaks can be ascribed to the quantum-confined Stark effect (QCSE) originating from an out-of-plane electric field associated with the ferroelectric polarization in AlScN.[39,40] Recent reports show that excitonic complexes in monolayer TMDs can exhibit an anomalous red shift in the presence of an external electric field, owing to strong confinement resulting from their atomic scale thickness [19,40]. We speculate that the observed arresting of the blue shift in the $X_T^+$ PL peak position at low temperatures is also associated to the QCSE. By fitting the broad PL spectrum with two Gaussian peaks (**see Supporting Information Figure S5**), we found spectral components associated with PL peak positions at ~1.64 and ~1.61 eV corresponding to positively charged bright ($X_T^+$) and dark trions or dark trion phonon replica ($X_{DT}^+$) , respectively. The peaks were assigned as per the previously reported and theoretically predicted binding energies in monolayer $WSe_2$.[41–43] In the $P_{up}$ polarized $WSe_2$/AlScN device (**Figure 3c**), most of the temperature-dependent excitonic features are the same, except that the broad PL emission at room temperature emerges from the negatively charged bright trions ($X_T^-$), with a binding energy ~28 meV below the $X^o$ PL peak. Additionally, we observe a red shift in the PL peak at 80 K when compared to the $P_{down}$ state device, which we again attribute to the QCSE. Furthermore, by fitting the broad PL emission, the spectral components with PL peak positions at 1.61 and 1.57 eV can be assigned to negatively charged bright trions ($X_T^-$) and dark trions or dark trion phonon replica ($X_{DT}^-$), with a binding energy ~26 meV below the $X_T^-$ peak position, respectively **(see supporting Information, Figure S6)**.[41–43]

We note that it is difficult to confirm if the ferroelectric polarization actually leads to the formation of dark trion/dark trion phonon replica or whether this effect is purely from the red shift of the bright trion PL peak combined with significant line-width broadening. Nevertheless, these fascinating excitonic characteristics, including both bright and dark trions, can have an important impact on the valley polarization dynamics in $WSe_2$. It has been reported that the formation of trions can significantly influence valley polarization properties. This is because they exhibit (i) ultralong valley lifetimes compared to the neutral excitons and (ii) reduced intervalley scattering, since trions involve an extra charge (i.e., an electron-hole pair and an excess charge carrier) for simultaneous large momentum transfer and spin flip.[34] Moreover, their net charge can also enable detection as well as manipulation of valley currents by external electric fields.[34,41]

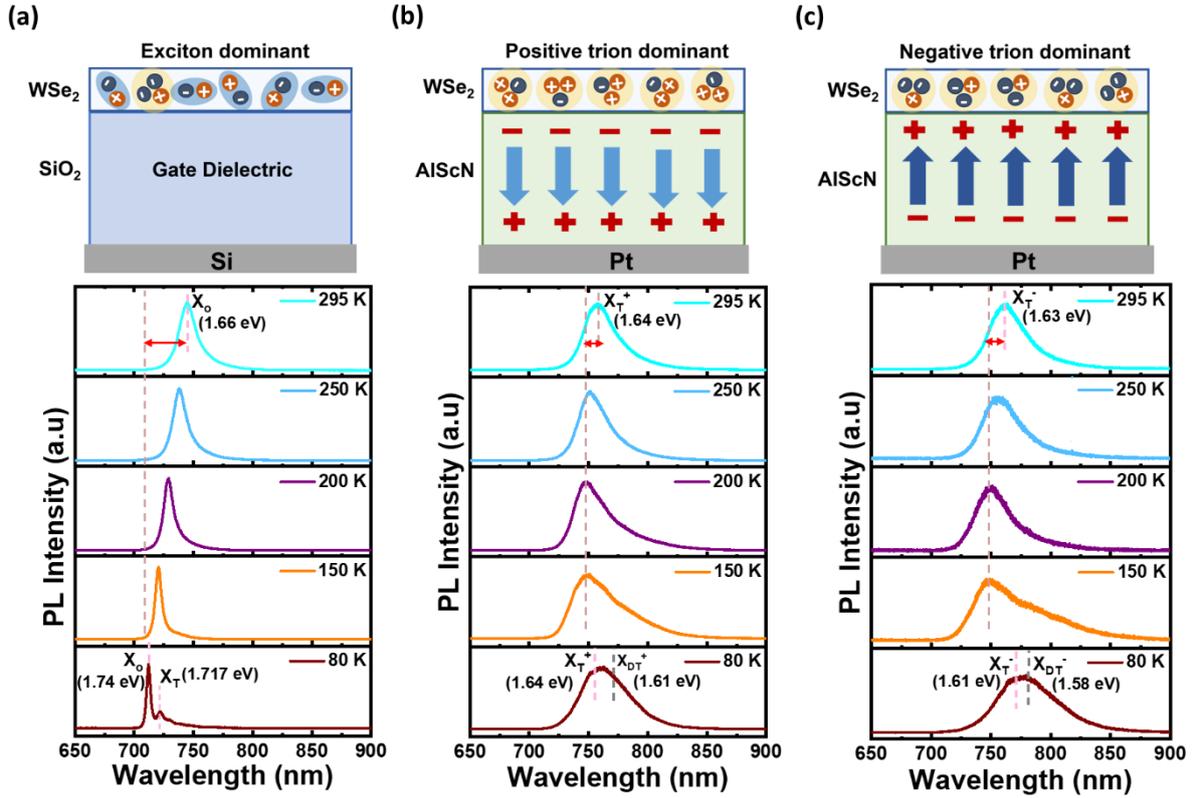

**Figure 3. Temperature dependent photoluminescence characteristics of monolayer WSe$_2$ exfoliated on** (a) non-ferroelectric SiO$_2$/Si substrate, (b) $P_{down}$ polarized and (c) $P_{up}$ polarized AlScN substrate. Top panels show dominant excitonic species in the devices. The WSe$_2$/SiO$_2$/Si devices is exciton dominant whereas, $P_{down}$ polarized the positively charged trions dominate the WSe$_2$/AlScN device due to the accumulation of excess of holes in WSe$_2$. And, $P_{up}$ polarized the negatively charged trions dominate the WSe$_2$/AlScN device due to the accumulation of excess electrons in the WSe$_2$ monolayer.

To study the valley polarization properties of WSe$_2$/AlScN FeFET device in two oppositely polarized states, we perform polarization-resolved PL measurements excited with a right-handed, circularly polarized (σ$^+$) 633 nm laser and detected with right-handed (σ$^+$) and left-handed (σ$^-$) helicity to measure the emission from *K* and *K'* valleys, respectively. **Figures 4a-b** display helicity-resolved PL spectra from a WSe$_2$/AlScN device polarized in the $P_{down}$ and $P_{up}$ states at 80 K. The degree of valley polarization (DVP) was calculated using the equation; $\eta = \frac{I(\sigma+) - I(\sigma-)}{I(\sigma+) + I(\sigma-)}$, where, *I(σ+)* and *I(σ-)* refers to the right- and left-handed circular polarization resolved PL intensity, respectively.[3] When the device is in the $P_{down}$ state, the $X_T^+$ shows a very low valley polarization degree of ~4%. While in the $P_{up}$ state, with increased negative charge carrier injection, the device exhibited a drastically higher DVP of ~27% in a representative WSe$_2$/AlScN device. This observation indicates that negatively charged trions have a greater impact on the valley polarization characteristics. The evolution of valley polarization as a function of temperature is also measured and is shown in **Figure 4c and Figures S7 – S8 (see supporting information)**. The temperature-dependent DVP shows a gradual decrease with an increase in the temperature. This is attributed to the increase in phonon-induced intervalley scattering leading to a change in the valley depolarization time. However, a valley polarization of ~7 % still persists at room temperature. This is likely because of the large polarization of the AlScN. To demonstrate the reproducibility of the valley polarization results with ferroelectric polarization switching, we designed and measured a 3 × 3 array of valleytronic devices in the $P_{up}$ state (**Figure 4e and Figure S9 see supporting information**). All the devices showed enhanced valley polarization, and a maximum DVP of ~ 27% was obtained among the nine devices measured. Furthermore, we measured the retention characteristics of the valley polarization in the $P_{up}$ state. A stable retention time of valley polarization up to 5400 secs (90 min.) was observed (**Figure 4d and Figure S10, see supporting information**). This result holds great promise for non-volatile optoelectronic memory that encodes both electrical conductance as well as polarization of the emission. The long retention time is particularly interesting and is mainly associated with the robustness of the ferroelectric polarization state. By controlling the laser heating effect and the robustness of the ferroelectric polarization, we anticipate that the retention time can be prolonged further. Similarly, devices fabricated using MOCVD-grown, large-area $WSe_2$ monolayers showed enhanced DVP in the $P_{up}$ state (**Figures S11 - S13, see supporting information**). Very few reports describe tuning of valley polarization characteristics, especially in monolayer TMDs. To our knowledge, ferroelectric tuning has never been reported. Thus, it is difficult to make a direct comparison. Nevertheless, we compare the tunability performance of our device with other tuning approaches, such as electrostatic gating and magnetic field tuning in **Figure 4f**.[18–20,44] It is evident that the ferroelectric gating approach in our device outperforms other tuning strategies. We find that valley polarization tuning of ~23% is obtained at relatively low voltages compared to electrostatic gating at higher voltages and high magnetic fields. Thus, our results indicate that ferroelectric polarization switching can be a potential knob to control valley polarization and achieve non-volatile valley addressable memory in monolayer $WSe_2$ at comparatively low power.

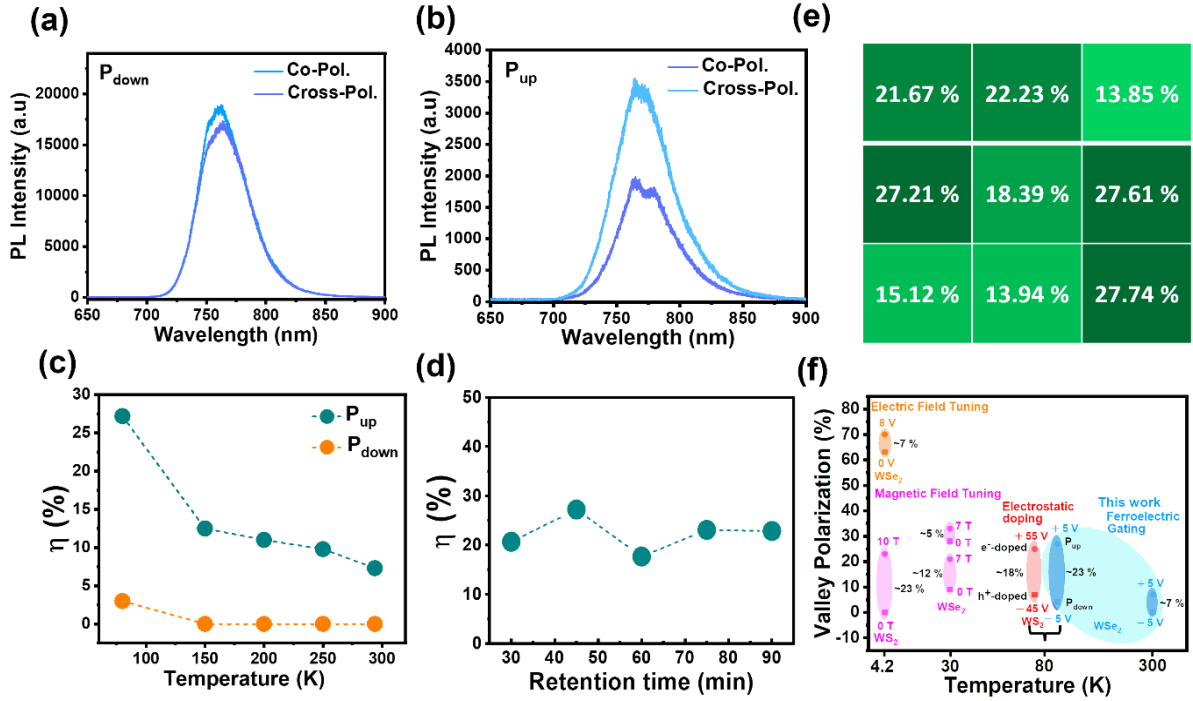

**Figure 4. Valley polarization characteristics of WSe$_2$/AlScN devices.** (a) Helicity resolved PL spectra from WSe$_2$/AlScN device in $P_{down}$ state (b) Helicity resolved PL spectra from WSe$_2$/AlScN device in $P_{up}$ state, at 80 K. (c) Temperature-dependent valley polarization characteristics of WSe$_2$/AlScN device in the $P_{up}$ and $P_{down}$ polarized states. (d) Retention characteristics of valley polarization in the $P_{up}$ polarized WSe$_2$/AlScN device (e) Valley polarization characteristics of a large 3 × 3 array of WSe$_2$/AlScN devices. (f) Comparison plot of previously reported values of valley polarization tuning in various monolayer TMD systems. The Orange dots denote tuning with an electric field. Magenta dots denote tuning with a magnetic field. Red dots denote tuning with electrostatic doping, and blue dots denote tuning with ferroelectric polarization switching in our work.[18–20,44]

The potential mechanism for the enhancement of valley polarization with ferroelectric polarization switching can be attributed to the formation of trions and strong QCSE-induced suppression of valley depolarization. For neutral excitons in monolayer TMDs, strong electron-hole (e-h) exchange interactions through the Maialle-Silva-Sham mechanism is the dominant valley depolarization process, which is relatively fast. This results in a short valley lifetime, on the order of few picoseconds,[14,15] as evidenced by the low DVP of excitons obtained in ML-WSe$_2$ exfoliated on the SiO$_2$/Si substrate (**Figure S14 see supporting information**). As the density of electron or hole injection increases with ferroelectric gating, excitons in WSe$_2$ can combine with extra charge carriers, leading to the formation of negatively or positively charged trions, respectively. For trions, valley depolarization is mainly determined by intervalley scattering because spin and momentum are mismatched. Previous studies on graphene and other TMD systems also show that the intervalley scattering rate is significantly suppressed with increased carrier doping.[41,45] Thus, valley depolarization of trions is expected to be much slower than observed for the excitons. As a result, trions show enhanced DVP.

To get more insight into the mechanism of valley polarization with ferroelectric gating mediated exciton to trion conversion *(X-T)*. The following valley resolved rate equations are utilized:[46]

$$\frac{dn_X^{K/K'}}{dt} = g^{K/K'} - n_X^{K/K'}\Gamma_X^r - n_X^{K/K'}\Gamma_{XT} + \left(n_X^{K'/K} - n_X^{K/K'}\right)\Gamma_X^{sk} \qquad (1.1)$$

$$\frac{dn_T^{K/K'}}{dt} = - n_T^{K/K'}\Gamma_T^r + n_X^{K/K'}\Gamma_{XT} + \left(n_T^{K'/K} - n_T^{K/K'}\right)\Gamma_T^{sk} \qquad (1.2)$$

here, $n_X^{K/K'}$ and $n_T^{K/K'}$ represent the density of excitons and trions in $K/K'$ valleys, $g^{K/K'}$ is the generation rate of excitons. $\Gamma_X^r$ and $\Gamma_T^r$ correspond to the recombination rate of excitons and trions, and $\Gamma_{XT}$ represents the exciton-to-trion (X-T) conversion rate. $\Gamma_X^{sk}$ and $\Gamma_T^{sk}$ represent intervalley scattering rates of excitons and trions, respectively. The solution of the above equations gives rise to DVP for excitons and trions, as follows:

$$\eta_X = \eta_o \frac{\Gamma_X^{ia}}{\Gamma_X^{ia} + \Gamma_X^{ir}} \qquad (2.1)$$

$$\eta_T = \eta_X \frac{\Gamma_T^{ia}}{\Gamma_T^{ia} + \Gamma_T^{ir}} \qquad (2.2)$$

Where $\eta_o$ is the initial DVP for excitons which is assumed to be 100 %, $\Gamma_X^{ia}$ & $\Gamma_T^{ia}$ stands for the intravalley decay rate of excitons and trions, and $\Gamma_X^{ir}$ & $\Gamma_T^{ir}$ denote the intervalley decay rate of excitons and trions, respectively. During the process of X-T conversion, the effective intravalley decay rate for excitons and trions can be described as follows:

$$\Gamma_X^{ia} = \Gamma_X^r + \Gamma_{XT} \qquad (3.1)$$

and,

$$\Gamma_T^{ia} = \Gamma_T^r \qquad (3.2)$$

As, the probability of trion to exciton conversion reduces at low temperatures, the term $\Gamma_{TX}$ is neglected in eqn. 3.2. For excitons, a fast decay of valley polarization is ascribed to the faster intravalley recombination, as well as the intervalley scattering rate, which is on the order of few ps.[14,15] Since trions mainly emerge from excitons, the valley polarization for trions depends on the exciton valley polarization. For example, under $\sigma^+$ circularly polarized light excitation, trions in the K valley are formed through the excitons generated in the K valley, while the trions in the K' valley are formed either by intervalley scattering of excitons from the K to the K' valley or by spin flipping of trions formed in the K valley. The probability of spin flipping is low due to the large momentum separation between valleys and, under the condition of fast X-T conversion rate in the K valley, the formation of trions in the K' valley and the intervalley decay rate of trions is extremely low, causing an imbalance of trions in K and K' valleys which leads to enhanced DVP for the trions. The above mechanism is schematically described in Figure 5a.

It is noteworthy that in our devices, positively and negatively charged trions exhibit an asymmetric valley polarization behaviour whereby negatively charged trions exhibit an enhanced valley polarization compared to the positively charged trions (**Figure 4a and 4b**). This phenomenon can be attributed to three reasons. First, the valley lifetime of negatively charged trions is longer than positively charged trions (**see supporting information Figure S15**).[42,47–51] Second, the strong quantum confined Stark effect in the $P_{up}$ state compared to the $P_{down}$ state leads to a higher suppression of electron-hole exchange interaction/intervalley scattering. As reported previously, the QCSE can effectively modify the overlap of electron-hole wave functions in monolayer TMDs, leading to a decrease in the intervalley scattering and spin relaxation rate.[19,52] Third, free electrons and holes possess different intervalley scattering mechanisms. For doped $WSe_2$ samples, the electron-hole recombination, which contains the valley information, is determined mainly by the minority charge carriers. i.e., electrons in the $P_{down}$ state device and holes in the $P_{up}$ state device.[51] In the $P_{up}$ state device, optically excited holes determine the valley excitonic emission in $WSe_2$ (**Figure 5b**), and the corresponding intervalley scattering (spin-flip) time for holes is quite long due to larger spin-splitting at the valence band ($\Delta E_{VB}$). Both theory and experimental studies have shown that the valley polarization lifetime of holes is on the order of nanoseconds.[53–55] Thus, the holes prefer to remain in the same valley as it is excited, leading to enhanced valley polarization. In the case of $P_{down}$ state device, the valley excitonic emission is determined by the optically excited electrons, which exhibit a very short intervalley scattering time (on the order of few picoseconds) due to smaller spin splitting in the conduction band ($\Delta E_{CB}$), resulting in less valley polarization. [56,57] Similar phenomena have been observed in other studies: the valley polarization is found to be suppressed in *p-doped* samples and enhanced in *n-doped* samples, and this has been attributed to a change in valley depolarization time.[20] Thus, we conclude that the long valley lifetime of the negatively charged trions and the strong coupling of the external electric field leads to enhanced valley polarization in our $P_{up}$ polarized WSe$_2$/AlScN device.

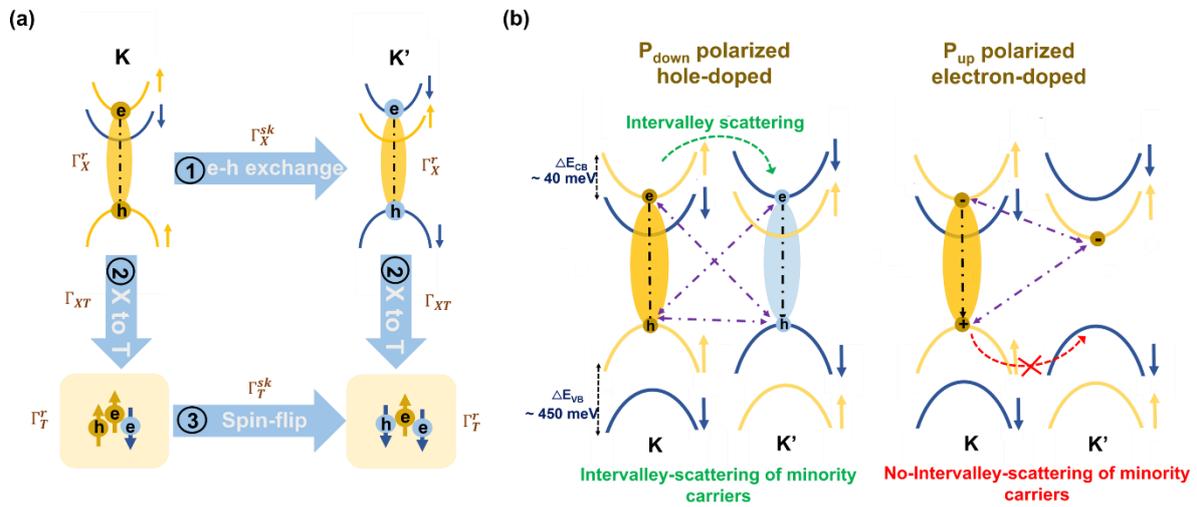

**Figure 5. Valley polarization mechanism in WSe$_2$/AlScN FeFET device.** (a) Schematic illustration of the formation of excitons and trions and their different relaxation pathways leading to valley depolarization. (b) Intervalley scattering and valley depolarization mechanism for negatively charged and positively charged trions.

**Conclusions**

In summary, we demonstrate a novel approach that uses ferroelectric materials to control valley polarization in monolayer TMDs. We show that the anticipated non-volatile electrical control of valley polarization could be realized experimentally by coupling a WSe$_2$ monolayer to a high $P_r$ ferroelectric such as AlScN. Enhanced, non-volatile valley polarization, robust even at room temperature, was demonstrated by controlling the polarization state of the AlScN. This approach will allow the control and manipulation of excited state and valley polarized phenomena in 2D and other quantum materials using ground-state ferroelectric polarization control. In principle, this largely unexplored aspect within heterostructure systems has the potential to pave the way for future ferro-valleytronic devices and sensors.

**Methods and Experimental Section**

**Sample Preparation and Device Fabrication.** *Mechanical exfoliation of µm-scale monolayer WSe$_2$.* WSe$_2$ monolayers were mechanically exfoliated from commercial bulk crystals (purchased from hq graphene Inc.) and transferred onto AlScN by the dry-transfer method.

The sample preparation process was carried out inside the glove box to avoid any sample degradation.

*Large area mechanical exfoliation and transfer of monolayer WSe$_2$*. Large area WSe$_2$ were exfoliated using gold tapes by following procedures in a previous study.[58] A gold tape was prepared by spin coating a polyvinylpyrrolidone (PVP) film on a 150 nm gold film on a SiO$_2$/Si substrate and then releasing a PVP/gold film by a thermal release tape. A released gold surface is attached to a freshly cleaved WSe$_2$ bulk crystal (HQ graphene) to produce a large area monolayer. An exfoliated monolayer was transferred onto a AlScN substrate.

*Synthesis of WSe$_2$ monolayer by MOCVD.* The growth of monolayer WSe$_2$ on 2" diameter c-plane sapphire was carried out in a metal-organic chemical vapor deposition (MOCVD) system equipped with a cold-wall horizontal reactor with an inductively heated graphite susceptor with gas-foil wafer rotation[59]. Tungsten hexacarbonyl (W(CO)$_6$) was used as the metal precursor while hydrogen selenide (H$_2$Se) was the chalcogen source with H$_2$ as the carrier gas. The W(CO)$_6$ powder was maintained at 30 °C and 400 Torr in a stainless-steel bubbler. The synthesis of monolayer WSe$_2$ is based on a multi-step process, consisting of nucleation, ripening, and lateral growth steps, which was described previously[60]. In general, the WSe$_2$ sample was nucleated for 30 sec at 850 °C, then ripened for 5min at 850 °C and 5 min at 1000 °C, and then grown for 20 min at 1000 °C, which gives rise to a coalesced monolayer WSe$_2$ across the entire 2" wafer. During the lateral growth, the tungsten flow rate was set as $3.8 \times 10^{-3}$ *sccm* and the chalcogen flow rate was set as 75 sccm while the reactor pressure was kept at 200 Torr. After growth, the substrate was cooled in H$_2$Se to 300 °C to inhibit the decomposition of the obtained monolayer WSe$_2$ film.

*Deposition of AlScN thin film*. AlScN were deposited on Pt (111)/Ti/SiO$_2$/Si wafers using Evatec CLUSTERLINE® 200 II pulsed DC sputtering system. Typically, 150 kHz pulsed DC co-sputtering with 20 sccm N$_2$ flow under $8.3 \times 10^{-4}$ mbar was used for the deposition. The temperature of the chamber was maintained at 350 ℃ during the deposition.

*Device fabrication*. The large-area exfoliated and MOCVD grown monolayer WSe$_2$ having a size of around mm$^2$ to cm$^2$ scale was transferred on a 1 cm$^2$-size AlScN wafer using the method explained above. Then, the WSe$_2$ exfoliated sample was coated using "PMMA A4 and PMMA A8" followed by source/drain (S/D) patterning via electron beam lithography (EBL). After developing it using MIBK, Ti/Au of 10 nm/50 nm layer was deposited using electron beam evaporation. Then to lift-off the deposited metal, the sample was soak in acetone for about 20 mins, and gently shaken, then rinsed with IPA. Next, to define the channel area, a second patterning using EBL was done after coating the same PMMA resists, followed by developing via MIBK. Finally, the open area after developing, the WSe$_2$ was etched using oxygen reactive ion etching (RIE).

**Characterizations.** *Electrical measurement of the WSe$_2$/AlScN FE-FET*. Electrical measurements were performed using the Lakeshore probe station using a Keithley 4200A semiconductor characterization system in air at room temperature. The SMU connection was used to obtain the DC I-V characteristics. The drain current was recorded with a 0.2 V step. For the polarization switching of the AlScN, the gate voltage of +10 or -10 V was applied while source is grounded.

*TEM/STEM sample preparation and measurements.* STEM/TEM cross-sectional samples of the devices were prepared by a Plasma-focused Ion Beam (TESCAN S8000X PFIB-SEM) system using the in-situ lift-out technique. The sample was coated with electron beam and ion beam deposition of Pt protection layers to prevent charging and heating effects (prevent damaging monolayer 2D material) during FIB milling. The final thinning with FIB was performed at 10 kV with 20 pA current to achieve a ~ 50 ± 5 nm sample thickness. TEM characterization and image acquisition were carried out on a JEOL F200 and a JEOL NEOARM operated at 200 kV accelerating voltage. The sample was orientated to the [001] zone axis for imaging. All of the captured TEM images were collected using Digital Micrograph software (DM, Gatan Inc., USA).

*Atomic force microscopy.* The layer thickness of WSe$_2$ monolayer transferred on AlScN substrate was determined using Omega Scope Smart SPM (AIST) atomic force microscopy.

*Optical measurements.* Micro-photoluminescence measurements were performed using Horiba LabRam HR Evolution confocal microscope. A 633 nm continuous wave laser was used as the excitation source. Temperature-dependent PL measurements in the temperature range of 80 -295 K with a cooling/heating rate of ~5 °C/min were performed using a liquid nitrogen cryostat. During the measurements the samples were kept in a Linkam stage under a vacuum level of ~10$^{-3}$ Torr. A 50x objective with NA ~ 0.9 was used for the sample excitation and the signal was collected by the same objective and electron multiplying charge coupled detector. For the circular polarization resolved PL measurements, a quarter wave plate was used to convert linearly polarized excitation light into circularly polarized light which was subsequently focused onto the sample. The signal is collected by the same quarter-wave plate. The right hand ($\sigma^+$) and left hand ($\sigma^-$) polarized PL signals is consequently converted to linear polarization and is selectively detected using a linear polarizer.

**Author Contributions**

S.S and D.J conceived the idea and designed the experiments. S.S carried out the small area mechanical exfoliation and B.K carried out large area mechanical exfoliation of monolayer WSe$_2$ on AlScN, respectively. N.T. and C.C. prepared wafer scale monolayer WSe$_2$ using MOCVD under the supervision of J.M.R. K.-H.K. fabricated valleytronic devices from the exfoliated and MOCVD grown samples and measured I-V characteristics. P.M carried out TEM/STEM measurements under the supervision of E.A.S. J.Z. performed sputtering of AlScN films under the supervision of R.H.O. S.S. performed photoluminescence spectroscopy, circularly polarized PL, AFM, Raman spectroscopy measurements with the help of K.J. S.S. analysed the data by discussion with all authors. S.S. and D.J wrote the paper. All authors discussed the results and commented on the manuscript.

**Acknowledgements:**

D.J. and K-H.K acknowledge primary funding support by the Air Force Office of Scientific Research (AFOSR) GHz-THz core program FA9550-23-1-0391. S.S. acknowledges support from the Fulbright-Nehru fellowship by United States-India Education Foundation (USIEF). S.S. also acknowledges partial support from the European Union's Horizon 2020 research and innovation programme under the Marie Skłodowska-Curie grant agreement No 899987. D.J. and K.J. acknowledge partial support from the Asian Office of Aerospace Research and Development (AOARD) of the AFOSR FA2386-20-1-4074 and FA2386-21-1-4063. K.J. acknowledges partial support from the Vagelos Institute of Energy Science and Technology Graduate Fellowship. Work by B.K. is supported by the U.S. Office of Naval Research (ONR) through grant N00014-20-1-2325 on Robust Photonic Materials with High-Order Topological Protection. R. O. acknowledges support from the NSF CAREER Award (1944248). N.T. acknowledges support from the National Science Foundation Graduate Research Fellowship Program under Grant No. DGE1255832. The MOCVD WSe$_2$ samples were provided by the 2D Crystal Consortium Materials Innovation Platform (2DCC-MIP) facility at Penn State which is funded by NSF under cooperative agreement DMR-2039351. Microfabrication and electron microscopy was performed at the Singh Center for Nanotechnology, supported by the NSF National Nanotechnology Coordinated Infrastructure Program (NNCI-1542153). The authors acknowledge the use of facilities and instrumentation supported by NSF University of Pennsylvania Materials Research Science and Engineering Center (MRSEC) (DMR-1720530).

# Supporting Information

**Non-Volatile Control of Valley Polarized Emission in 2D WSe$_2$-AlScN Heterostructures**

Simrjit Singh[a,b], Kwan-Ho Kim[a], Kiyoung Jo[a], Pariasadat Musavigharavi[a,c,d], Bumho Kim[e], Jeffrey Zheng[c], Nicholas Trainor[f], Chen Chen[g], Joan M. Redwing[f,g], Eric A Stach[c], Roy H Olsson III[a], Deep Jariwala[a]

[a]Department of Electrical and Systems Engineering, University of Pennsylvania, Philadelphia, PA, USA.

[b]Department of Applied Physics and Science Education, Eindhoven University of Technology, Eindhoven, The Netherlands.

[c]Department of Materials Science and Engineering, University of Pennsylvania, Philadelphia, PA, USA.

[d] Department of Materials Science and Engineering, University of Central Florida, Orlando, FL, USA.

[e]Department of Physics and Astronomy, University of Pennsylvania, Philadelphia, PA, USA

[f]Department of Materials Science and Engineering, Pennsylvania State University, University Park, PA, USA.

[g]2D Crystal Consortium Materials Innovation Platform, Materials Research Institute, Pennsylvania State University, University Park, PA, USA.

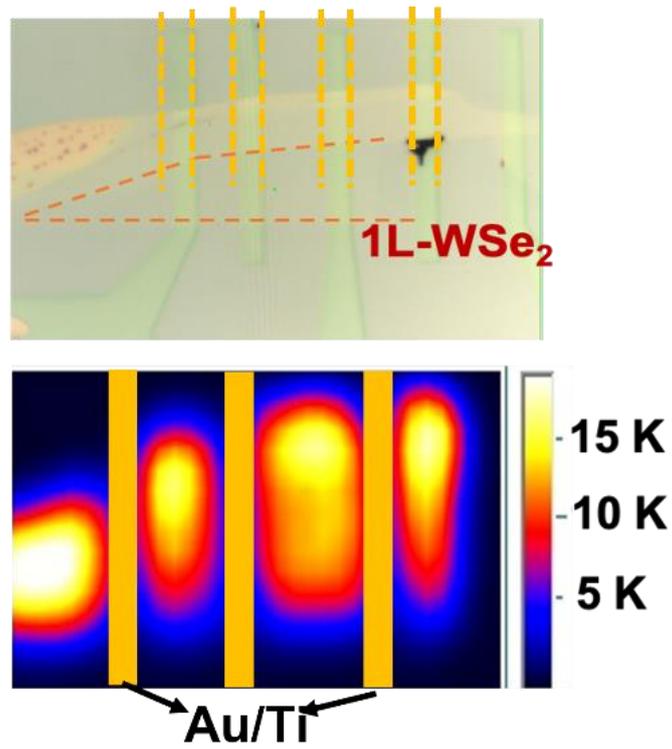

**Figure S1:** Optical image and corresponding PL mapping image of mechanically exfoliated monolayer WSe$_2$/AlScN FeFET devices.

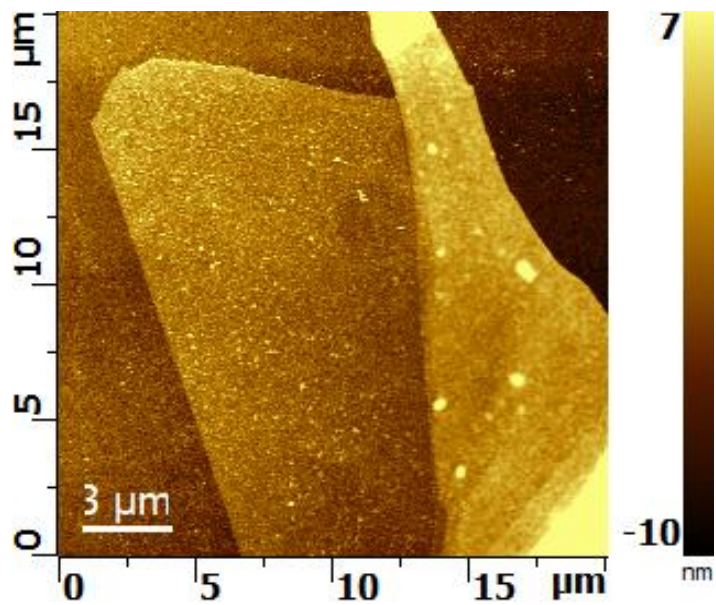

**Figure S2:** AFM image of exfoliated monolayer WSe$_2$ on AlScN substrate.

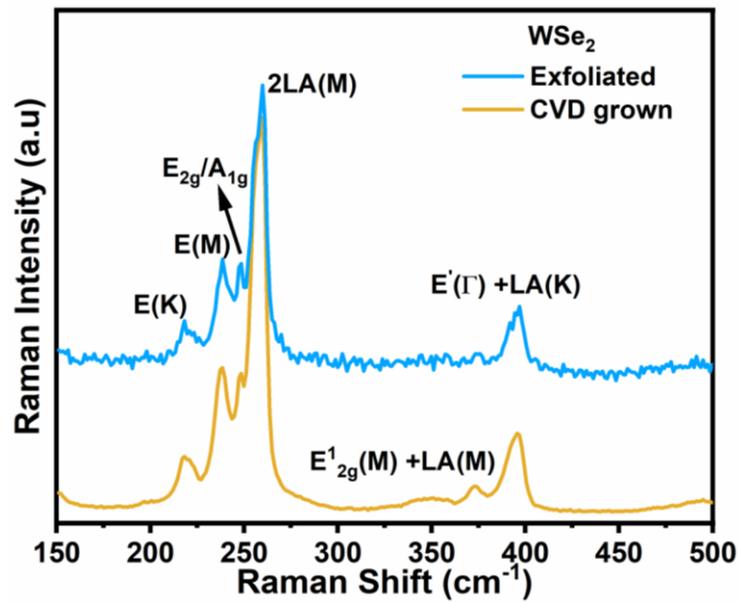

**Figure S3:** : Raman Spectra of exfoliated and MOCVD grown monolayer WSe$_2$ on AlScN substrate.

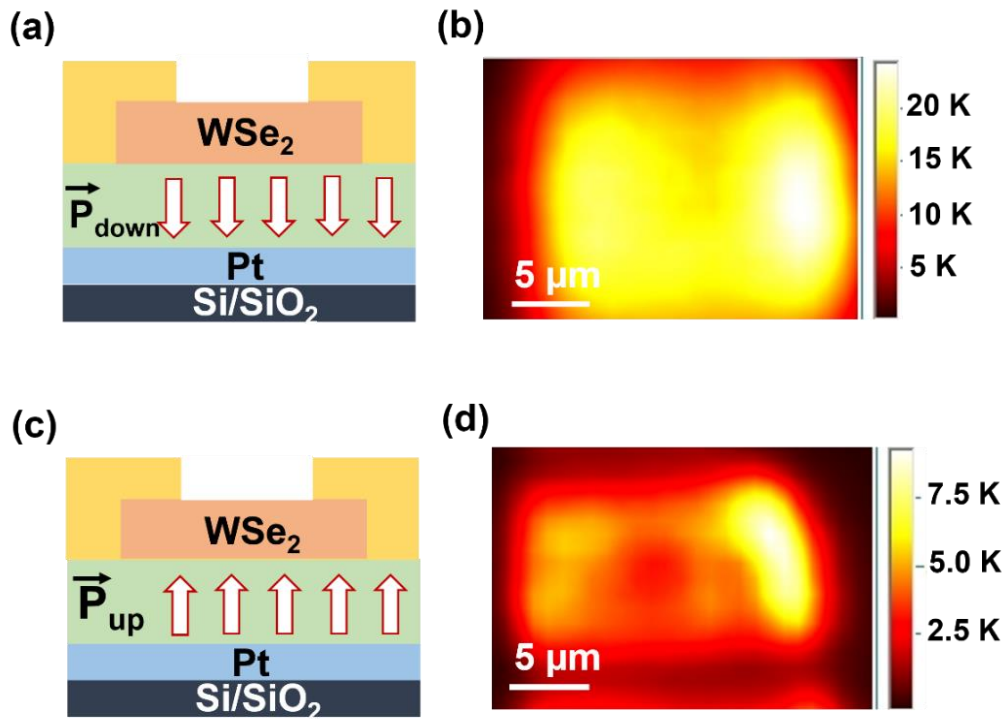

**Figure S4:** PL intensity maps from mechanically exfoliated WSe$_2$/AlScN FeFET device in (a,b) downward polarized state and (c,d) upward polarized state.

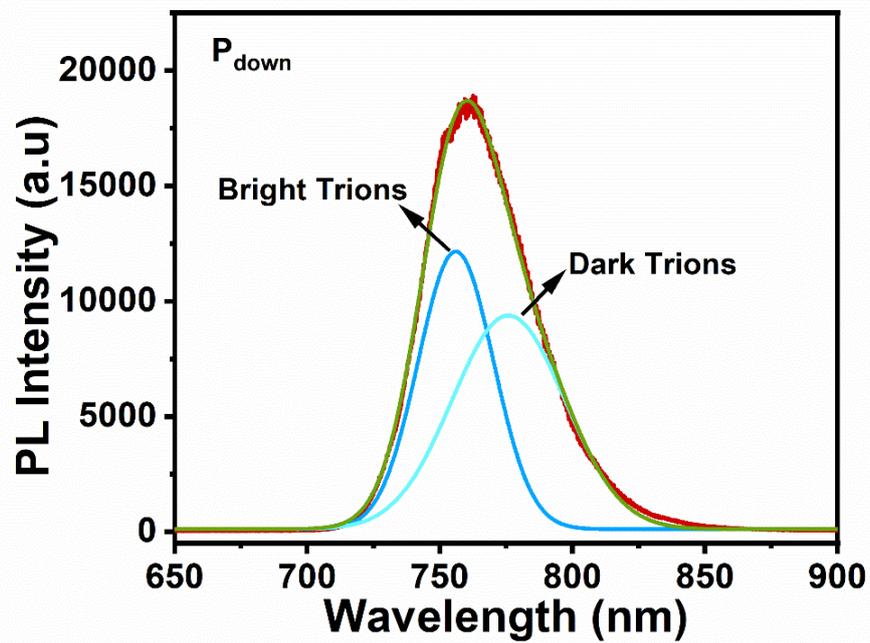

**Figure S5:** PL spectra of $P_{down}$ state WSe$_2$/AlScN device at 80 K. The PL peak fitting shows contribution from the bright and dark positively charged trions.

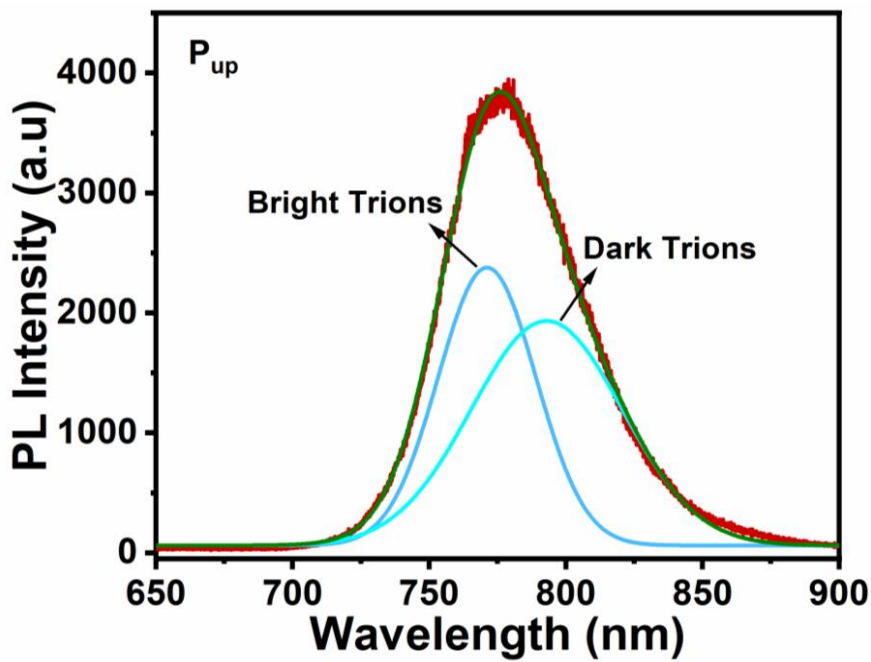

**Figure S6:** PL spectra of $P_{up}$ state WSe$_2$/AlScN device at 80 K. The PL peak fitting shows contribution from the bright and dark negatively charged trions.

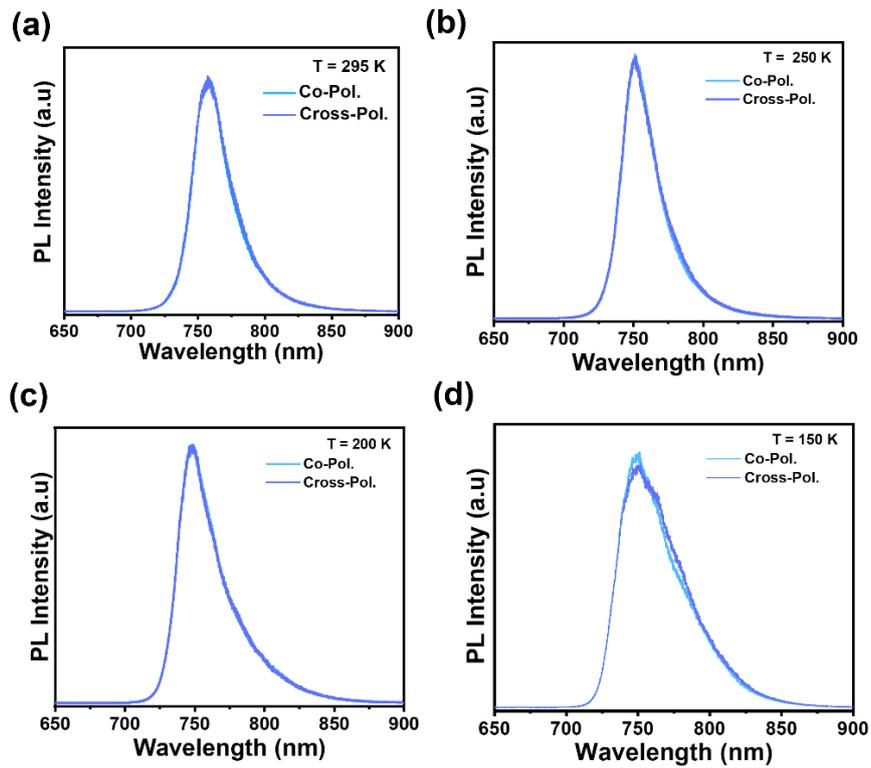

**Figure S7:** Temperature dependent helicity resolved photoluminescence measurements on mechanically exfoliated WSe$_2$/AlScN device in the $P_{down}$ state.

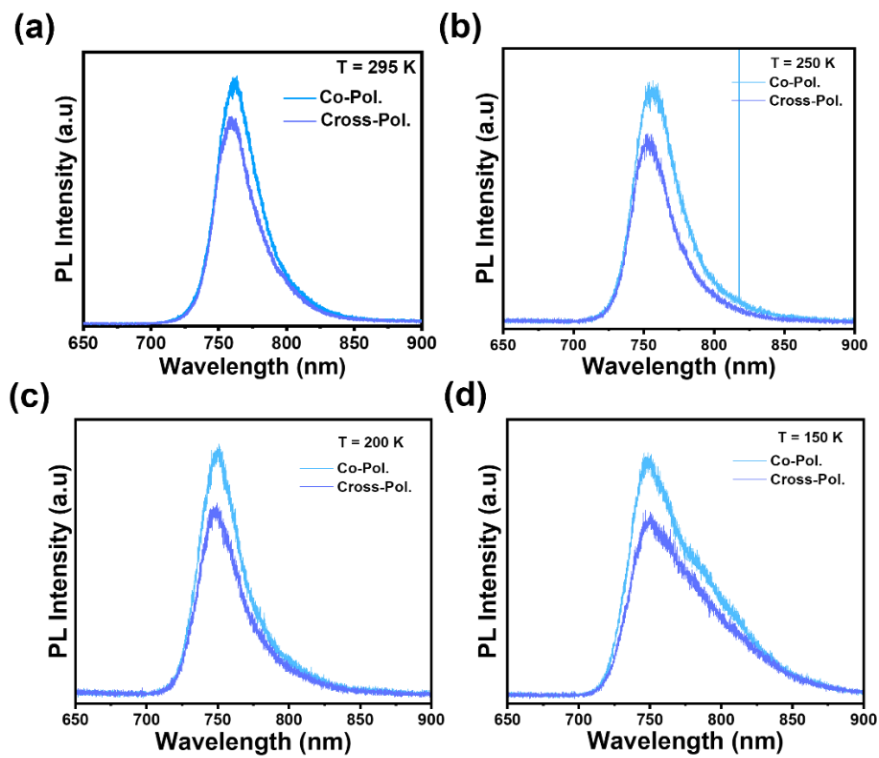

**Figure S8:** Temperature dependent helicity resolved photoluminescence measurements on mechanically exfoliated WSe$_2$/AlScN device in the *P$_{up}$* state.

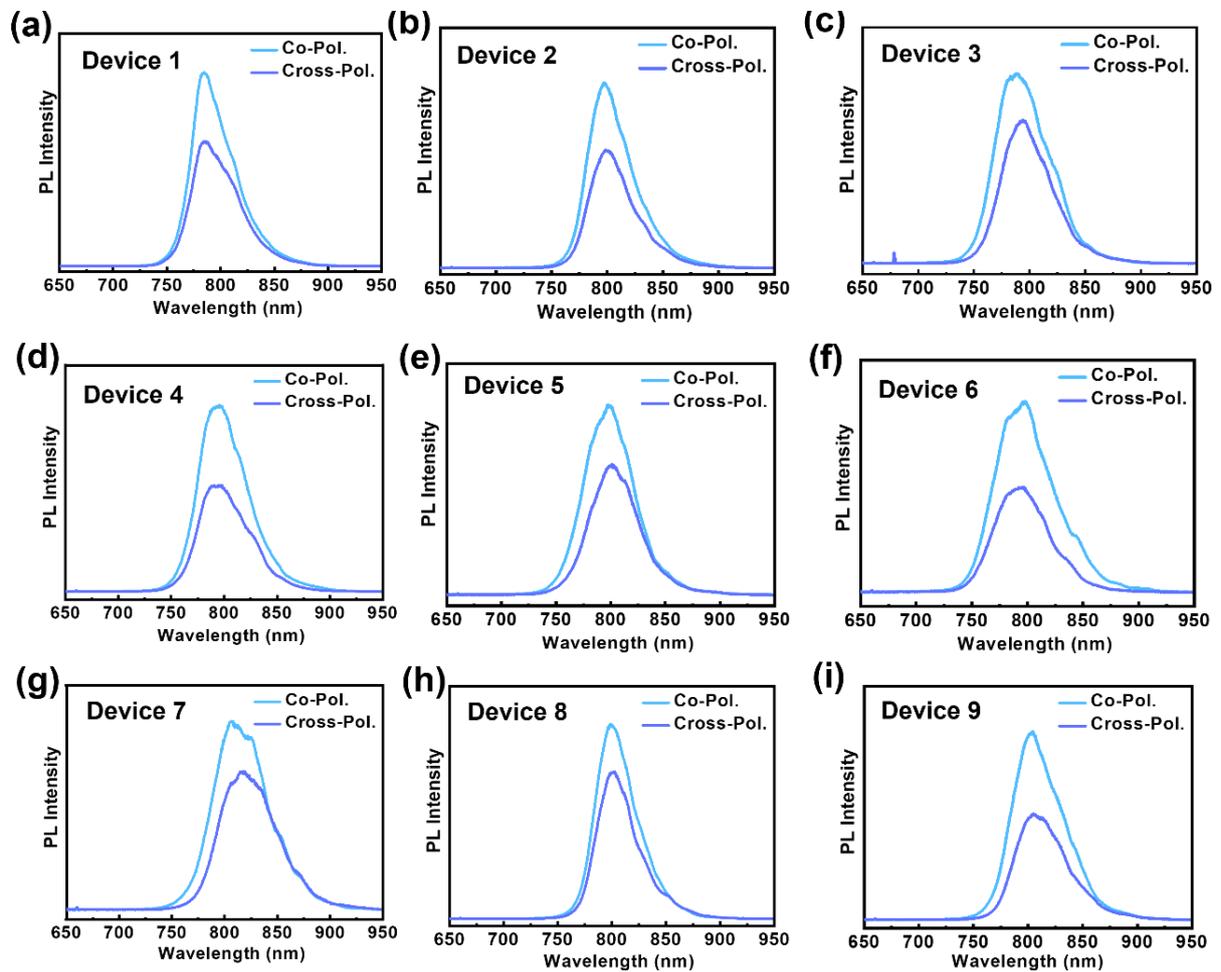

**Figure S9:** Helicity resolved photoluminescence measurements from large area 3×3 array of WSe$_2$/AlScN devices in the P$_{up}$ state at 80 K.

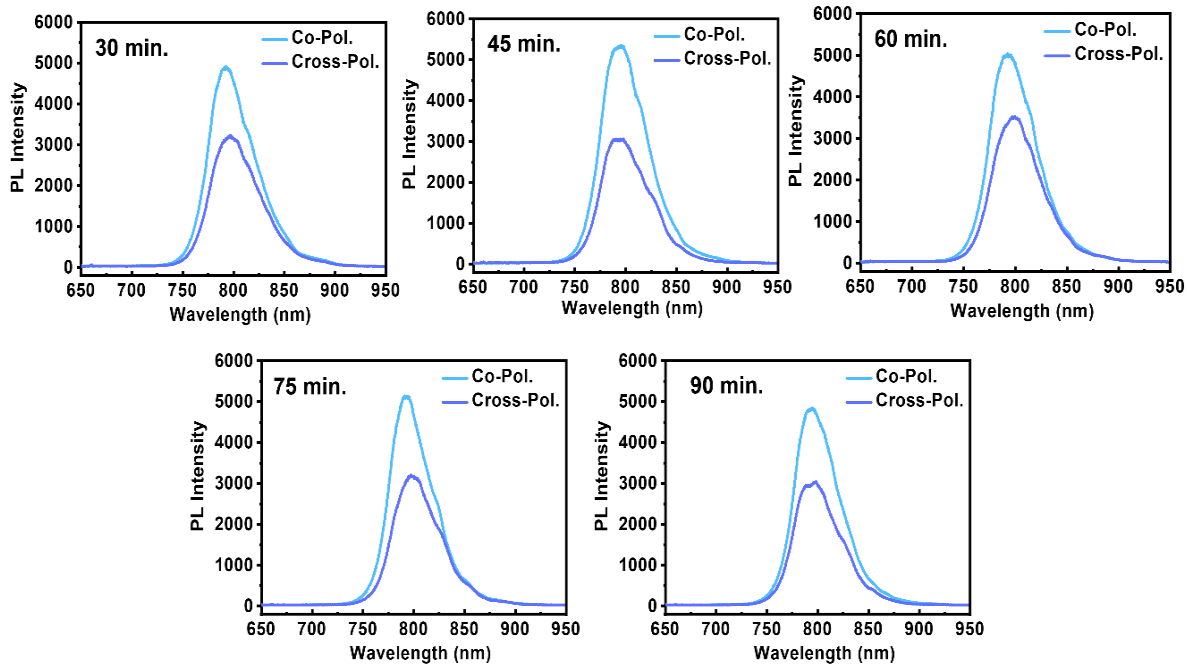

**Figure S10:** Time dependent helicity resolved photoluminescence measurements from a representative WSe$_2$/AlScN device in the $P_{up}$ state at 80 K.

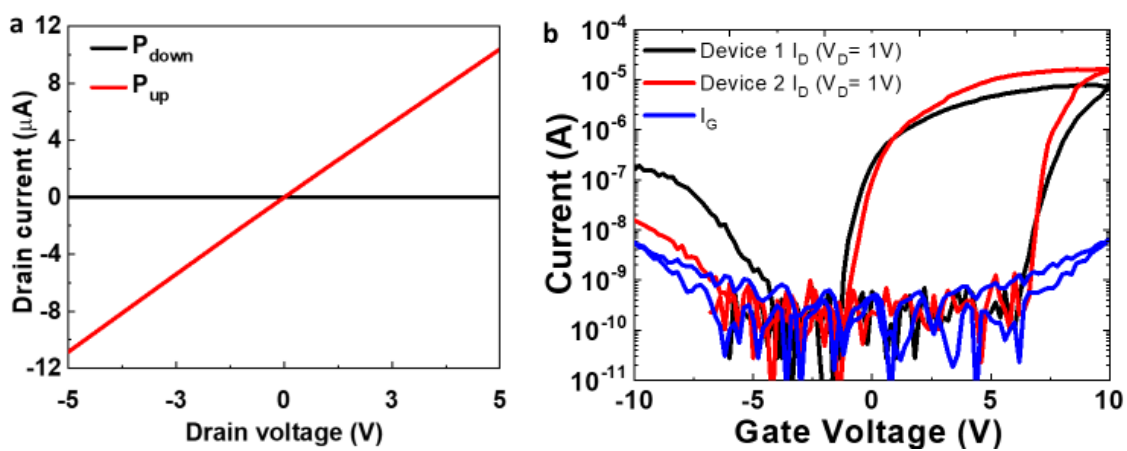

**Figure S11**: a. Output (Drain current, $I_D$) characteristics of WSe$_2$/AlScN Fe-FETs in the $P_{up}$ and $P_{down}$ states at gate voltage ($V_G$) =0 V showing large difference in resistance. b. Transfer characteristics of WSe$_2$/AlScN Fe-FETs showing counterclockwise hysteresis loops (typically of n-type Fe-FETs) and large modulation of $I_D$ with $V_G$.

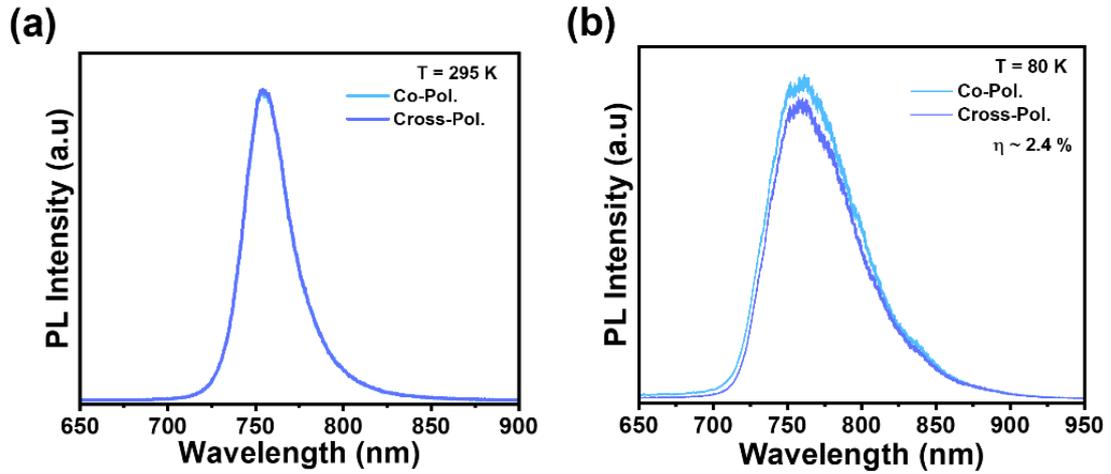

**Figure S12:** Helicity resolved photoluminescence measurements on MOCVD grown-WSe$_2$/AlScN device in the *P$_{down}$* state measured at (a) 295 K and (b) 80 K.

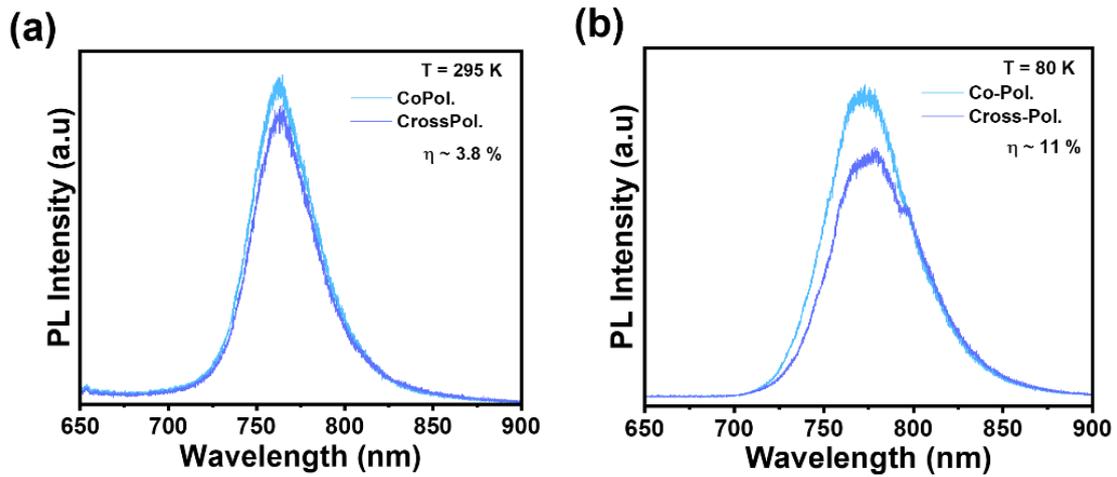

**Figure S13:** Helicity resolved photoluminescence measurements on MOCVD grown-WSe$_2$/AlScN device in the *P$_{up}$* state measured at (a) 295 K and (b) 80 K.

The degree of valley polarization (η) is found to be reduced in devices fabricated with MOCVD WSe$_2$ monolayer compared to exfoliated material which can be due to the presence of inversion domains in epitaxially grown monolayers.[1]

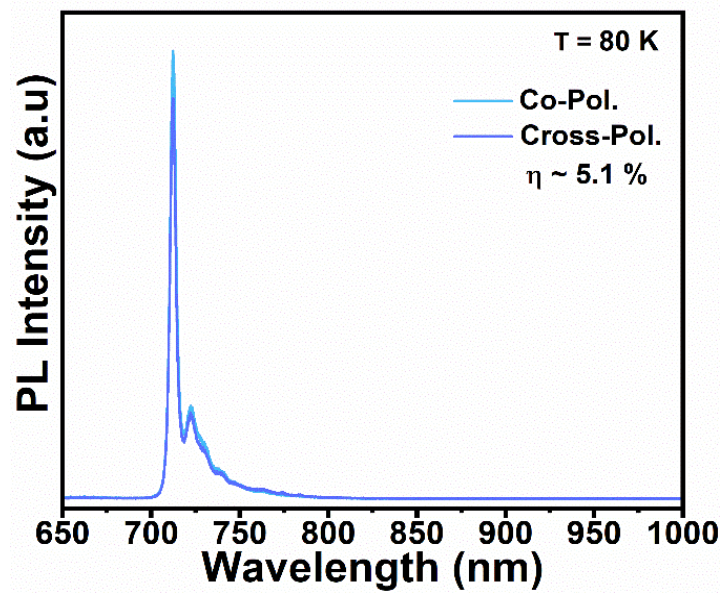

**Figure S14:** Helicity resolved photoluminescence measurements on WSe$_2$/SiO$_2$/Si device in at 80 K.

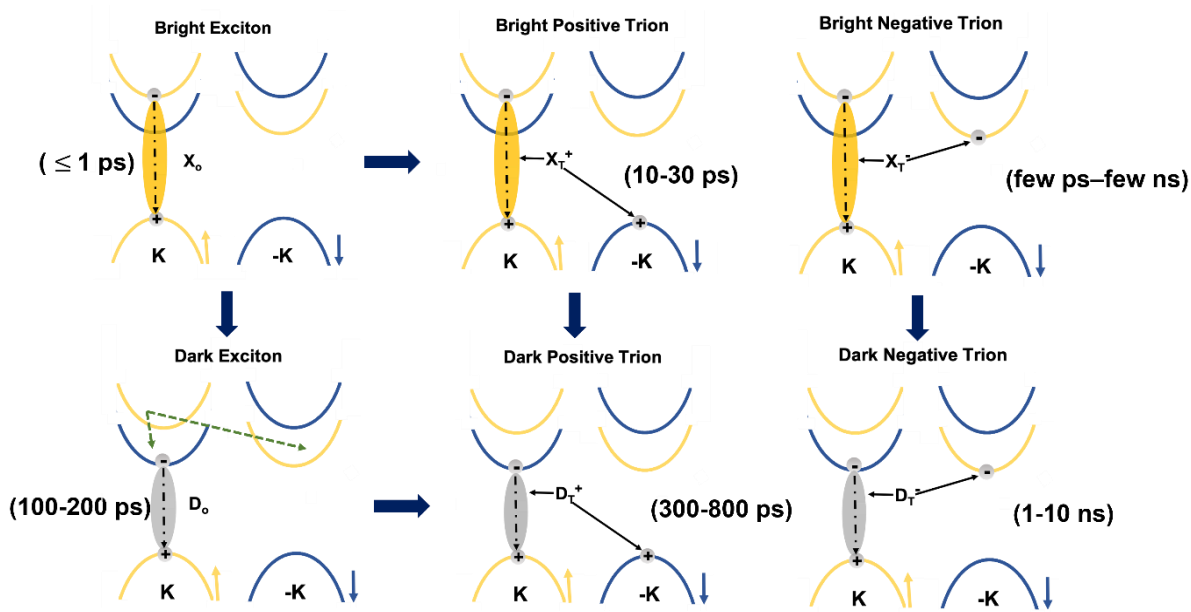

**Figure S15:** Band structure and formation of different excitonic complexes in WSe$_2$ with their corresponding lifetimes reported in the literature. (a) bright excitons, (b) dark excitons, (c) Positively charged bright trions and (d) Positively charged dark trions (e) Negatively charged bright trions and (f) Negatively charged dark trions.[2-11]

From literature precedent it can be concluded that there are mainly two plausible mechanisms for valley depolarization i.e., (i) electron-hole exchange interactions and (ii) intervalley scattering rate.[2–4] For neutral excitons, electron-hole exchange interaction is the dominant valley depolarization mechanism whereas, in the case of trions valley depolarization is mainly dominated by the intervalley scattering rate. Notably, the valley depolarization mechanism and timescale are substantially different for neutral excitons and charged excitonic species i.e., both bright and dark trions.[5–8] The valley depolarization time of bright trions is reported to be on the order of 10-70 ps. On the other hand, valley depolarization time of dark trions is reported to be two order of magnitude longer (~1.3 ns) than the bright trions.[9–11] **Figure S15** display schematic illustration of different excitonic complexes with their respective valley depolarization lifetimes reported for $WSe_2$. Monolayer $WSe_2$ possess a unique band structure in which the valence and conduction bands split into two sub bands with opposite spins due to strong spin-orbit coupling. Thus, it can host bright as well as dark excitonic complexes; recombining electron-hole pair in the same valley having same spin orientation is termed as bright exciton whereas, recombining electron-hole pairs with antiparallel spin results in spin-forbidden dark excitons which gives rise to strongly quenched radiative recombination.[9,10] With ferroelectric polarization switching, bright and dark excitons can be coupled to the additional charge carriers leading to the formation of charged bright and dark trions. A high density of positive and negative charges doped in $WSe_2$ results in positively and negatively charged bright/dark trions, respectively. Thus, the formation of bright and spin forbidden dark trions in our samples play an important role in enhancing the valley polarization owing to suppressed valley relaxation time.